\documentclass[%
 preprint,
 amsmath,amssymb,
 aps,
numbers=noenddot;
]{revtex4-1}

\usepackage{graphicx}% Include figure files
\usepackage{dcolumn}% Align table columns on decimal point
\usepackage{bm}
\usepackage{mathtools}
\usepackage{breqn}
\usepackage{physics}
\usepackage{caption}
\usepackage{subcaption}
\usepackage{subcaption}
\DeclareCaptionLabelSeparator{none}{ }
\begin{document}

%\preprint{APS/123-QED}

\title{	
Closed Form Solutions For The Quantum LRS Bianchi IX and VIII  Wheeler DeWitt Equations For An Arbitrary Hartle-Hawking Ordering Parameter. 
}% Force line breaks with \\

\author{Daniel Berkowitz}
 \altaffiliation{Physics Department, Yale University.\\ daniel.berkowitz@yale.edu \\ This work is in memory of my parents, Susan Orchan Berkowitz, and Jonathan Mark Berkowitz}%Lines break automatically or can be forced with \\

\date{\today}% It is always \today, today,
             %  but any date may be explicitly specified

\begin{abstract}
In this paper we present the quantum LRS Bianchi IX models or, as they may be better known, the Taub-NUT models in addition to the quantum LRS Bianchi VIII models, both coupled to a stiff matter source. We solve their Wheeler DeWitt equations for an arbitrary Hartle-Hawking ordering parameter using separation of variables. Afterwards we construct a superposition of their respective wave functions and discuss some of the interesting qualitative properties they possess.

\end{abstract}

\pacs{Valid PACS appear here}% PACS, the Physics and Astronomy
                             % Classification Scheme.
%\keywords{Suggested keywords}%Use showkeys class option if keyword
                              %display desired
\maketitle

%\tableofcontents

\section{\label{sec:level1}INTRODUCTION}
  
The Bianchi IX models possess a rich history and have been studied in a plethora of contexts. Investigations into the Bianchi IX models began with the works of Misner\cite{misner1969mixmaster}, Ryan\cite{ryan1972oscillatory}, and Belinskii, Khalatnikov and Lifshitz (BKL)\cite{belinskii1970oscillatory}\cite{belinskii1978asymptotically}. What partly made these models so appealing was that they shared the 3-sphere spatial topology of the k=1 FLRW models, which previously were widely believed to be a good approximation for our physical universe, until more precise cosmological observations \cite{de2002multiple} showed otherwise. Furthermore the equations which govern the dynamics of its mini-superspace variables, which we will take to be the Misner variables $(\alpha,\beta_+,\beta_-)$ \cite{misner1969mixmaster}\cite{misner1969quantum} appear to admit chaotic solutions \cite{barrow1982chaotic} \cite{chernoff1983chaos} \cite{cornish1997mixmaster}. The standard quantum diagonalized Bianchi IX models were first studied by Misner \cite{misner1969quantum}, and later by Moncrief-Ryan \cite{moncrief1991amplitude}, among many others. Our focus in this work will be the quantum LRS Bianchi IX and VIII models. The LRS Bianchi IX models are also known as the Taub-NUT models and have a very interesting history behind them \cite{taub1951empty} \cite{newman1963empty}. A separable closed form solution for the quantum LRS Bianchi IX models(TAUB model) when Hartle-Hawking ordering\cite{hartle1983wave} $B=0$ was found by Martinez and Ryan\cite{martinez1983exact} in the early 80s. In comparison the Bianchi VIII models and their associated LRS models have not been studied as thoroughly. However some important results pertaining to the Bianchi VIII models have been compiled \cite{banerjee1984spatially} \cite{lorenz1980exact} \cite{karagiorgos2019quantum}. 

In this paper we will further extend the results originally obtained by Martinez and Ryan\cite{martinez1983exact} by solving the symmetry reduced Wheeler DeWitt equations of both the LRS Bianchi IX and VIII models coupled to a stiff matter source for any Hartle-Hawking ordering parameter using separation of variables. Afterwards we will analyze the interesting features that a superposition of our solutions possess. 

\section{\label{sec:level1} Bianchi IX and VIII LRS Wave Functions}

Classically the cosmologies we will be study possess the following metrics 

\begin{equation}
\begin{aligned}
&d s^{2}=-d t^{2}+e^{2 \alpha(t)}\left(e^{2 \beta(t)}\right)_{a b} \omega^{a} \omega^{b}\\& \left(e^{2 \beta(t)}\right)_{a b}=\operatorname{diag}\left(e^{2 \beta_{+}\left(t\right)}, e^{2 \beta_{+}\left(t\right)}, e^{-4 \beta_{+}\left(t\right)}\right).
\end{aligned}
\end{equation}

The $\omega^{i}$ terms are one forms defined on the spatial hypersurface of each Bianchi cosmology and obey $ d \omega^{i}=\frac{1}{2} C_{j k}^{i} \omega^{j} \wedge \omega^{k}$ where $C_{j k}^{i}$ are the structure
constants of the invariance Lie group associated with each particular class of Bianchi models. For the Bianchi IX and VIII models they are

\begin{equation}
\begin{aligned} \omega^{1} &=d x-k \sinh (k y) d z \\ \omega^{2} &=\cos (x) d y-\sin (x) \cosh (k y) d z \\ \omega^{3} &=\sin (x) d y+\cos (x) \cosh (k y) d z, \end{aligned}
\end{equation}

where k=i corresponds to the Bianchi IX models and k=1 corresponds to the Bianchi VIII models.

In Misner variables the Wheeler DeWitt equations for the LRS Bianchi IX and VIII models are the ordinary Bianchi IX and VIII Wheeler Dewitt equations presented in \cite{obregon1996psi} when $\beta_-$ = 0, and the $\beta_-$ degree of freedom in their kinetic terms are removed. Using the methodologies presented in \cite{ryan1972hamiltonian}\cite{pazos2000aplicacion} a perfect fluid matter source in a comoving frame is added with the equation of state $P= \rho$. This results in the following Wheeler Dewitt equations

\begin{equation}
\frac{\partial^2 \Psi}{\partial \alpha^2}-B\frac{\partial \Psi}{\partial \alpha}-\frac{\partial^2 \Psi}{\partial \beta_+^{2}}+\left( \frac{e^{4 \alpha-8 \beta_+}}{3} \left(1\pm4 e^{6 \beta_+} \right)+384\pi GM\right)\Psi=0,
\end{equation}
where $384\pi GM$ is an integration constant for our stiff matter source which has been normalized to keep our results aesthetically in line with previously established solutions to the Wheeler DeWitt equation \cite{aguero2007noncommutative}.
The plus sign in (3) corresponds to the LRS Bianchi VIII models, and the minus sign is for the LRS Bianchi IX models, and B is the Hartle-Hawking \cite{hartle1983wave} ordering parameter. To solve equation (3) we will use separation of variables, and our first step is to perform the following change of variables suggested by \cite{chernoff1983chaos}

\begin{equation}
\begin{aligned}
\xi=4\alpha-8\beta_+\\ \kappa=4\alpha-2\beta_+.
\end{aligned}
\end{equation}

After expressing the kinetic terms in our new variables (4) we obtain the following separable partial differential equation

\begin{equation}
12\frac{\partial^2 \Psi}{\partial \kappa^2}-4B\frac{\partial \Psi}{\partial \xi}-4B\frac{\partial \Psi}{\partial \kappa}-48\frac{\partial^2 \Psi}{\partial \xi^{2}}+\left(\frac{e^{\xi}}{3} \pm \frac{4 e^{\kappa}}{3}+384\pi GM\right)\Psi=0.
\end{equation}

Using separation of variables we first set$\Psi$ = $\Xi(\xi)K(\kappa)$, and insert this into equation (5) resulting in 

\begin{equation}
12\Xi(\xi)\frac{d^2 K}{d\kappa^2}-4BK(\kappa)\frac{d \Xi}{d \xi}-4B\Xi(\xi)\frac{d K}{d\kappa}-48K(\kappa)\frac{d^2 \Xi}{d \xi^{2}}+\left(\frac{e^{\xi}}{3} \pm \frac{4 e^{\kappa}}{3}+C_{0}\right)\Xi(\xi)K(\kappa)=0
\end{equation}

where $C_{0}=384\pi GM$. Afterwards we divide equation (6) by $\Xi(\xi)K(\kappa)$ and bring all of the terms which are solely functions of $\kappa$ and $\xi$ to opposing sides. This yields the following ordinary differential equations 

\begin{equation}
\frac{12}{K(\kappa)}\frac{d^2 K}{d\kappa^2}-\frac{4B}{\Xi(\xi)}\frac{d \Xi}{d \xi}-\frac{4B}{K(\kappa)}\frac{d K}{d\kappa}-\frac{48}{\Xi(\xi)}\frac{d^2 \Xi}{d \xi^{2}}+\left(\frac{e^{\xi}}{3} \pm \frac{4 e^{\kappa}}{3}+C_{0}\right)=0
\end{equation}

\begin{equation}
\frac{12}{K(\kappa)}\frac{d^2 K}{d\kappa^2}-\frac{4B}{K(\kappa)}\frac{d K}{d\kappa} \pm \frac{4 e^{\kappa}}{3}+C_{0}=\frac{48}{\Xi(\xi)}\frac{d^2 \Xi}{d \xi^{2}}+\frac{4B}{\Xi(\xi)}\frac{d \Xi}{d \xi}-\frac{e^{\xi}}{3}.
\end{equation}

From separation of variables we can set both of our functions of $\kappa$ and $\xi$ in equation (8) equal to a constant which will give us two differential equations

\begin{equation}
\begin{aligned}
\frac{12}{K(\kappa)}\frac{d^2 K}{d\kappa^2}-\frac{4B}{K(\kappa)}\frac{d K}{d\kappa} \pm \frac{4 e^{\kappa}}{3}+C_{0}=-\gamma^2\\\frac{48}{\Xi(\xi)}\frac{d^2 \Xi}{d \xi^{2}}+\frac{4B}{\Xi(\xi)}\frac{d \Xi}{d \xi}-\frac{e^{\xi}}{3}=-\gamma^2.
\end{aligned}
\end{equation}

We choose our separation constant, $-\gamma^2$ to be negative so that our resulting wave functions qualitatively possess the characteristics of a wave as opposed to an exponential. For both the LRS Bianch IX and VIII models the solutions of these differential equations are Bessel functions. Our separable wave functions of the universe for the LRS Bianch IX and VIII models are respectively 

\begin{equation}
\begin{aligned}
&\Psi_{IX \hspace{1mm} \gamma}=K_{IX}(\kappa)\Xi_{IX}(\xi)\\& K_{IX}(\kappa)=\left(c_2 e^{\frac{B \kappa}{6}} I_{\pm\frac{1}{3} \sqrt{B^2-3 \gamma^2-3 C_{0}}}\left(\frac{2 \sqrt{e^{\kappa}}}{3}\right)+c_1 e^{\frac{B \kappa}{6}} K_{\pm\frac{1}{3} \sqrt{B^2-3
   \gamma^2-3 C_{0}}}\left(\frac{2 \sqrt{e^{\kappa}}}{3}\right)\right)\\ & \Xi_{IX}(\xi)=\left(c_4 e^{-\frac{B \xi}{24}} I_{\pm\frac{1}{12} \sqrt{B^2-12
   \gamma^2}}\left(\frac{\sqrt{e^{\xi}}}{6}\right)+c_3 e^{-\frac{B \xi}{24}} K_{\pm\frac{1}{12}
   \sqrt{B^2-12 \gamma^2}}\left(\frac{\sqrt{e^\xi}}{6}\right)\right)
\end{aligned}
\end{equation}

\begin{equation}
\begin{aligned}
&\Psi_{VIII \hspace{1mm} \gamma}=K_{VIII}(\kappa)\Xi_{VIII}(\xi)\\& K_{VIII}(\kappa)=\left(c_2 e^{\frac{B \kappa}{6}} J_{\pm\frac{1}{3} \sqrt{B^2-3 \gamma^2-3 C_{0}}}\left(\frac{2 \sqrt{e^{\kappa}}}{3}\right)+c_1 e^{\frac{B \kappa}{6}} Y_{\pm\frac{1}{3} \sqrt{B^2-3
   \gamma^2-3 C_{0}}}\left(\frac{2 \sqrt{e^{\kappa}}}{3}\right)\right)\\ & \Xi_{VIII}(\xi)=\left(c_4 e^{-\frac{B \xi}{24}} I_{\pm\frac{1}{12} \sqrt{B^2-12
   \gamma^2}}\left(\frac{\sqrt{e^{\xi}}}{6}\right)+c_3 e^{-\frac{B \xi}{24}} K_{\pm\frac{1}{12}
   \sqrt{B^2-12 \gamma^2}}\left(\frac{\sqrt{e^\xi}}{6}\right)\right)
\end{aligned}
\end{equation}
where $c_{i}$ can be any real or complex numbers. When $B$ and $C_{0}$ vanish (10) reduces to the solution originally found in \cite{martinez1983exact}.  We will return later to how the choice of the Hartle-Hawking ordering parameter B non-trivially affects the nature of our wave functions. 
   
Because the Wheeler DeWitt equation is linear one can construct additional wave functions which exhibit interesting properties by constructing a superposition $\int^\infty_{-\infty} e^{-b\left(g-\gamma\right)^2}\Psi_{\gamma}d\gamma$, where b and g are any real or complex number. The only requirements we will impose on our wave functions is that they are smooth and globally defined. 

Before we plot our wave functions we need to discuss the so called problem of time. We can see the problem of time manifest itself if we relate the Wheeler DeWitt equation, $\hat{\mathcal{H}}_{\perp} \Psi=0$, to the time dependent Schr$\text{\" o}$dinger equation

\begin{equation}
\begin{aligned}
& i \hbar \frac{\partial \Psi}{\partial t}=N \hat{\mathcal{H}}_{\perp} \Psi \\ &
\frac{\partial \Psi}{\partial t}=0.
\end{aligned}
\end{equation}

It appears that time "vanishes" in Wheeler DeWitt quantum cosmology. A way around this for our purposes is to denote one of the Misner variables to be our clock. A good clock increases monotonically. Out of the variables we can choose from $\alpha$, which corresponds to the scale factor of the LRS models we are studying, is the best candidate for our clock and will be for practical purposes our "time" \cite{dewitt1967quantum} parameter. Using the aforementioned internal clock interpretation, our plots will be 2D representations of the wave functions. However, due to the apparent difficulty in constructing a dynamical unitary operator from the symmetry reduced Wheeler DeWitt equation, $|\psi\left(\alpha,\beta_+\right)|^{2}$ is not conserved in $\alpha$, thus we cannot assign a simple probabilistic interpretation to our wave functions. In general it appears that there are fundamental difficulties in constructing a unitary operator corresponding to a quantum anisotropic cosmology \cite{alvarenga2003troubles}. 

In this paper in addition to treating $\alpha$ as our time parameter, we will also treat it on an equal footing to our anisotropic $\beta_+$ variable and plot 3D representations of our wave functions. The author hopes that despite the interpretational issues surrounding $\Psi$, plotting our wave functions in these two manners will shed some light on what they are qualitatively trying to tell us about the quantum evolution of the universes they respectively describe.

For the LRS Bianchi IX and VIII models as the reader can verify the "Bessel I" functions approach infinity when either $\alpha$ approaches $\infty$ or $\beta_+$ approaches $-\infty$. This behavior complicates interpreting our results, and thus we will exclude the "Bessel I" functions while plotting our wave functions by setting $c_{2}=c_{4}=0$ in (10) and $c_{4}=0$ in (11). Furthermore, as the readers can verify for themselves, when the indices $\sqrt{B^2-3 \gamma^2-3 C_ {0}} $ and $\sqrt{B^2-12 \gamma^2}$ that characterize our Bessel functions are not purely imaginary they behave like exponentials as opposed to waves which heavily complicates interpreting them. Thus we will include a lower bound cut off in our integral for our superpositions so that the aforementioned indices are purely imaginary, which will facilitate us in interpreting our results. 

Non exotic matter sources $C_ {0} \geq 0$ in addition to real values of our integration constant $\gamma$ contribute to our coefficients being imaginary. However, real values of our Hartle-Hawking parameter B can allow our coefficients to be real for certain values of $\gamma$ and $C_ {0} $. This alone does not pose any complications because we are free to choose our integration constant $\gamma$ so that our coefficients are always imaginary, resulting in our wave functions having the aesthetic behavior that is expected of them such as having identifiable peaks and troughs. However the terms $e^{\frac{B \kappa}{6}}$ and $e^{-\frac{B \xi}{24}}$ when $B \ne 0$ non-trivially affects the nature of our wave functions by causing their peaks and troughs to either grow or decay exponentially depending upon the value of $B$, and which Misner variables $\alpha$ and $\beta_+$ are being varied. This behavior does not necessarily make our wave functions when $B \ne 0$ non-physical. However, for the purposes of this paper due to the complications introduced by having $B \ne0$ we will only analyze the case when $B=0$. This peculiar behavior when $B \ne 0$ can merely be a result of our separable solutions and does not in of itself show that this behavior is a defining feature of the quantum LRS Bianchi IX and VIII models. Furthermore a rigorous definition for the norm of Wheeler DeWitt wave functions can potentially be found which results in this peculiar behavior vanishing.

For the quantum LRS Bianchi IX models, through numerical integration we plot the following wave function $\abs{\int^\infty_{1} e^{-\gamma^2}\Psi_{IX \hspace{1mm} \gamma}d\gamma}^{2}$ where B=0, and $c_{1}$=1, $c_{2}$=0, $c_{3}$=1 and $c_{4}$=0, for the range of $C_{0}$ listed below 

\begin{figure}[!ht]
\begin{minipage}[c]{0.4\linewidth}
\includegraphics[scale=.15]{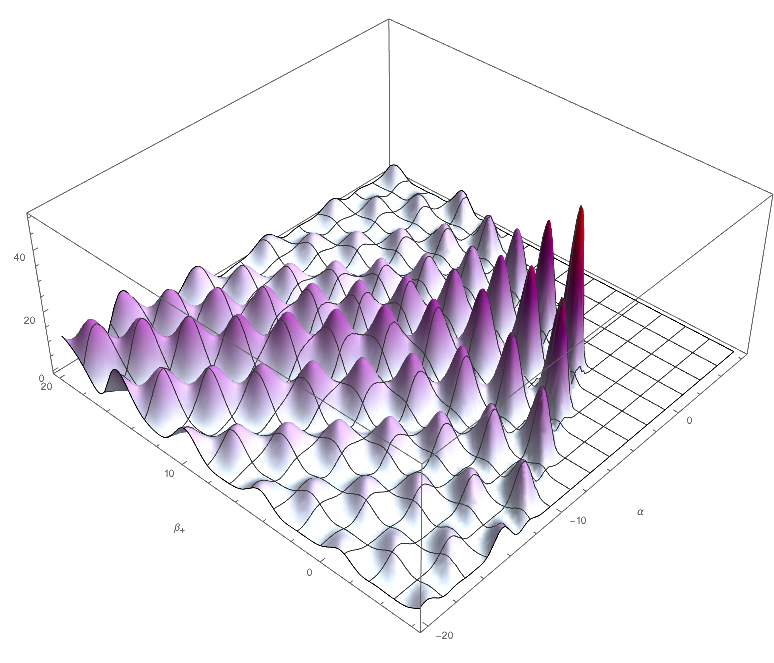}
\caption{$C_{0}$=0}
\end{minipage}
\hfill
\begin{minipage}[c]{0.4\linewidth}
\includegraphics[scale=.15]{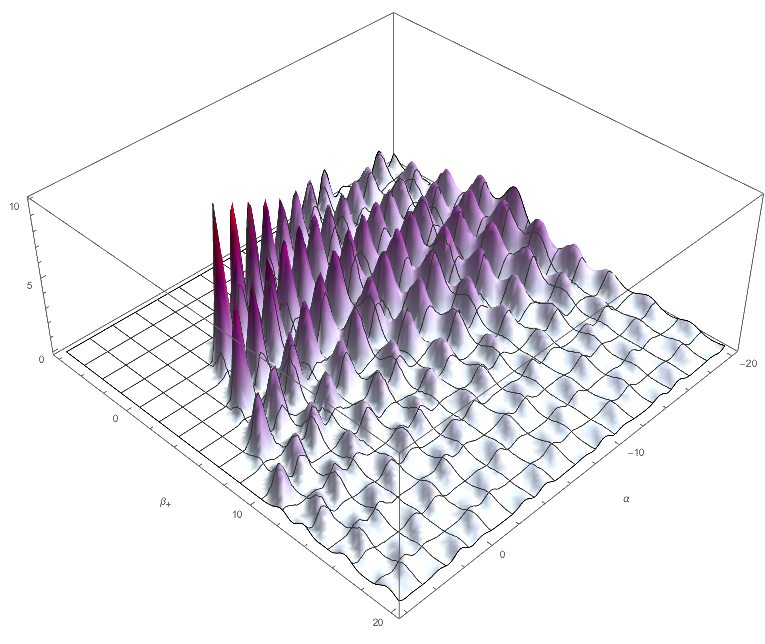}
\caption{$C_{0}$=2}
\end{minipage}%
\end{figure}

\begin{figure}[!ht]
\begin{minipage}[c]{0.4\linewidth}
\includegraphics[scale=.15]{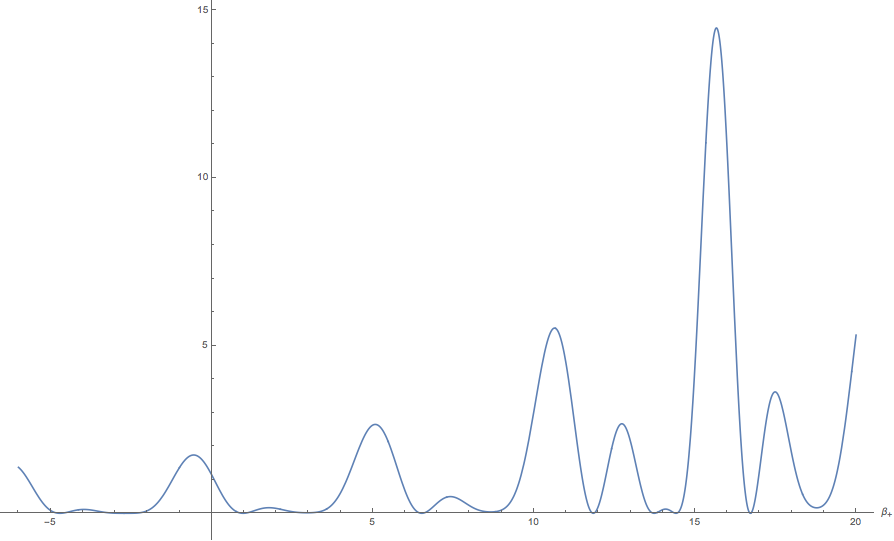}
\caption{$C_{0}$=0 \hspace{1mm} $\alpha$=-20}
\end{minipage}
\hfill
\begin{minipage}[c]{0.4\linewidth}
\includegraphics[scale=.15]{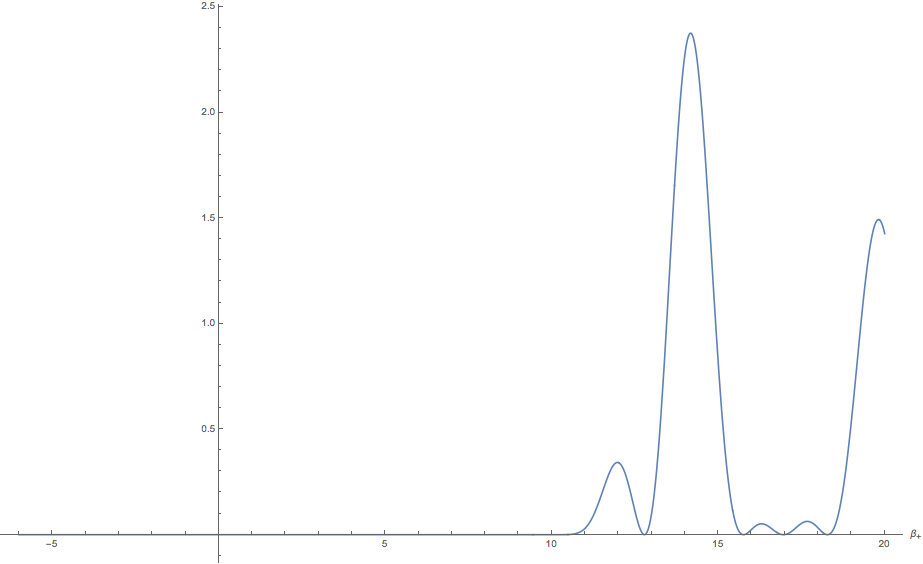}
\caption{$C_{0}$=0 \hspace{1mm} $\alpha$=6}
\end{minipage}%
\end{figure}

\begin{figure}[!ht]
\begin{minipage}[c]{0.4\linewidth}
\includegraphics[scale=.15]{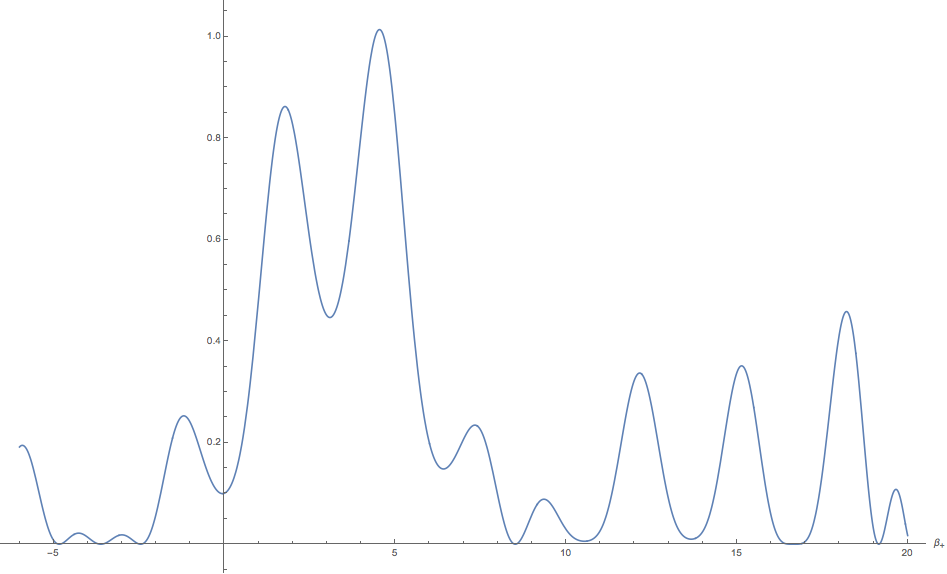}
\caption{$C_{0}$=2 \hspace{1mm} $\alpha$=-20}
\end{minipage}
\hfill
\begin{minipage}[c]{0.4\linewidth}
\includegraphics[scale=.15]{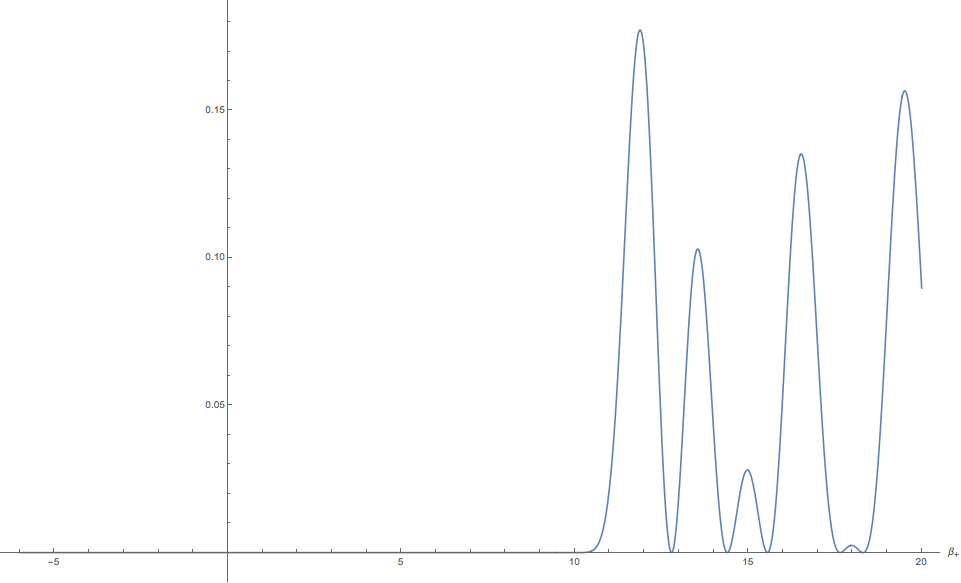}
\caption{$C_{0}$=2 \hspace{1mm} $\alpha$=6}
\end{minipage}%
\end{figure}

Likewise for the Bianchi VIII models, through numerical integration we plot the following wave function $\abs{\int^\infty_{1} e^{-\gamma^2}\Psi_{VIII \hspace{1mm} \gamma}d\gamma}^{2}$ where B=0, and $c_{1}$=1, $c_{2}$=1, $c_{3}$=1 and $c_{4}$=0, for the range of $C_{0}$ listed below 

\begin{figure}[!ht]
\begin{minipage}[c]{0.4\linewidth}
\includegraphics[scale=.15]{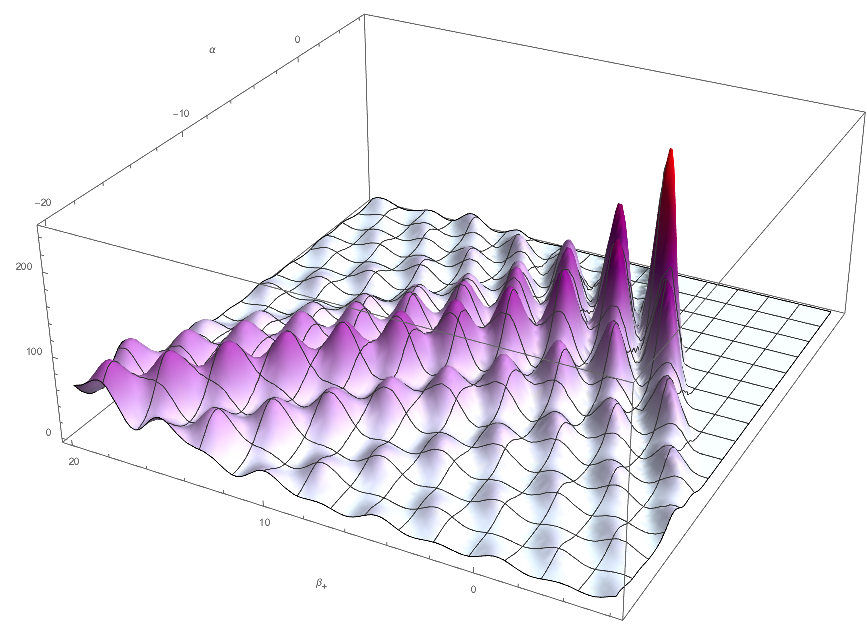}
\caption{$C_{0}$=0}
\end{minipage}
\hfill
\begin{minipage}[c]{0.4\linewidth}
\includegraphics[scale=.13]{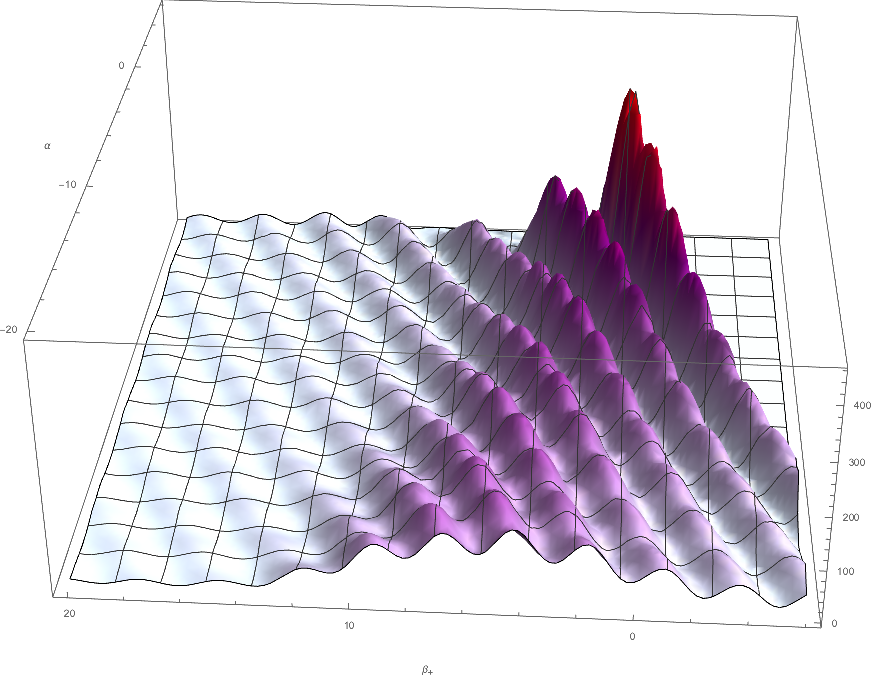}
\caption{$C_{0}$=2}
\end{minipage}%
\end{figure}

\begin{figure}[!ht]
\begin{minipage}[c]{0.4\linewidth}
\includegraphics[scale=.15]{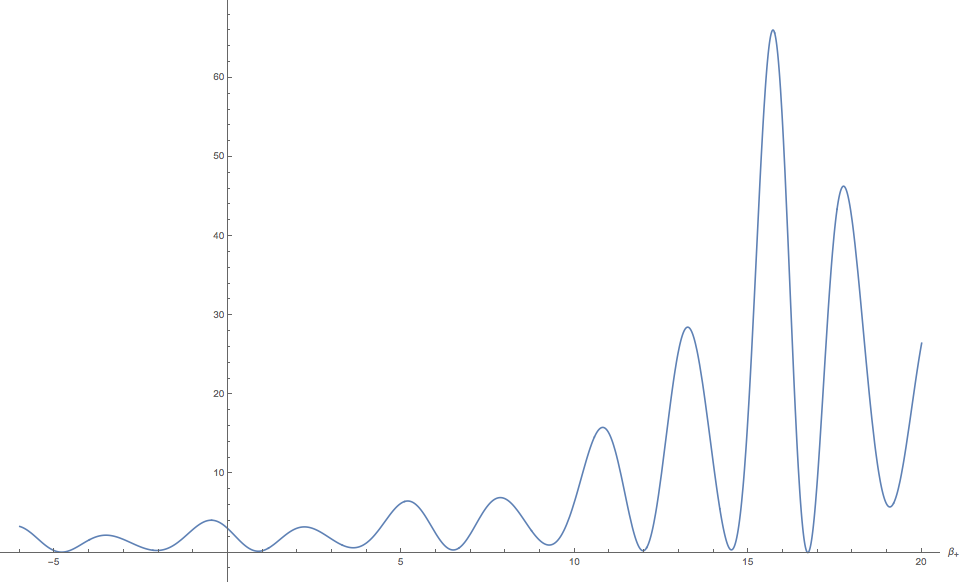}
\caption{$C_{0}$=0 \hspace{1mm} $\alpha$=-20}
\end{minipage}
\hfill
\begin{minipage}[c]{0.4\linewidth}
\includegraphics[scale=.15]{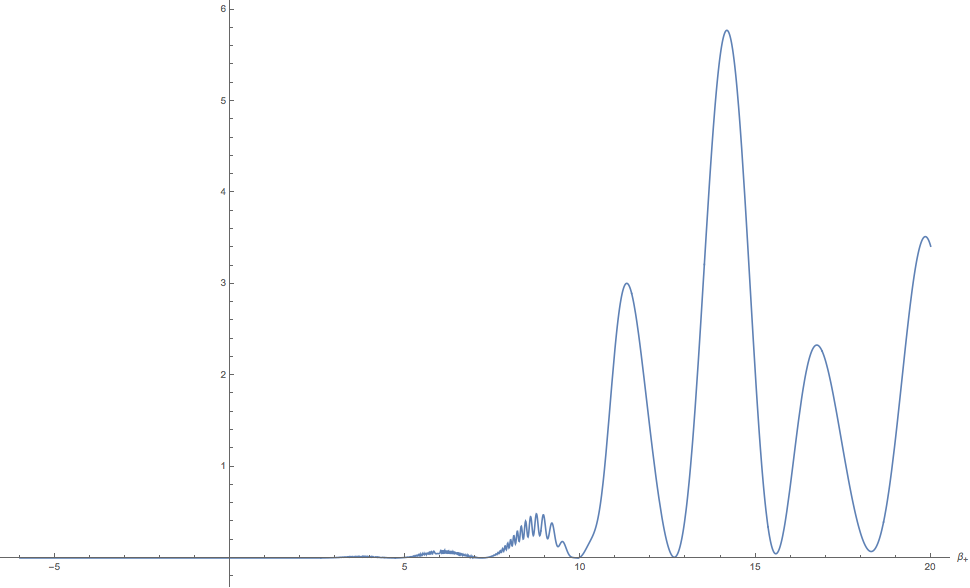}
\caption{$C_{0}$=0 \hspace{1mm} $\alpha$=6}
\end{minipage}%
\end{figure}

\begin{figure}[!ht]
\begin{minipage}[c]{0.4\linewidth}
\includegraphics[scale=.15]{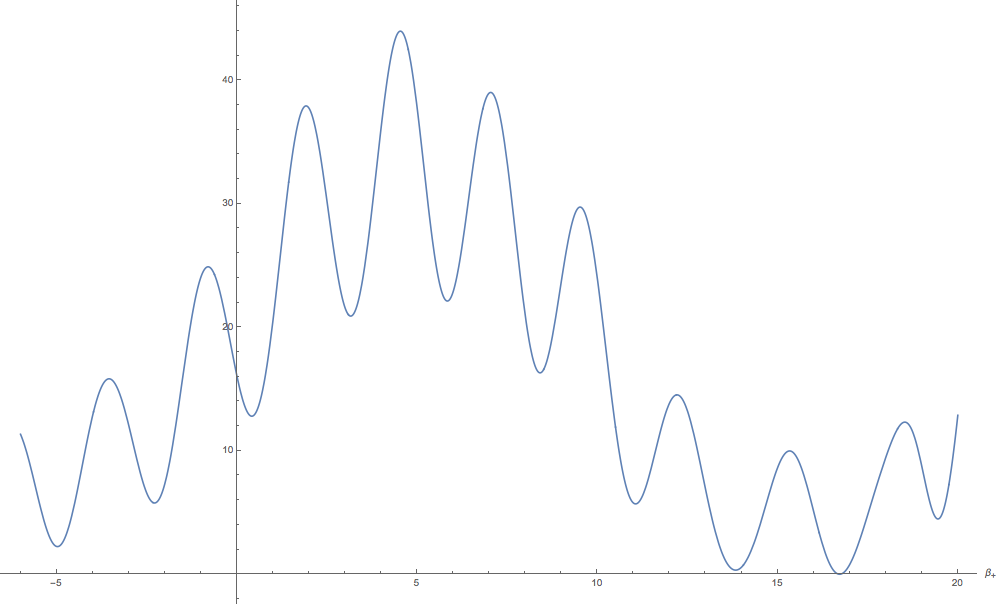}
\caption{$C_{0}$=2 \hspace{1mm} $\alpha$=-20}
\end{minipage}
\hfill
\begin{minipage}[c]{0.4\linewidth}
\includegraphics[scale=.15]{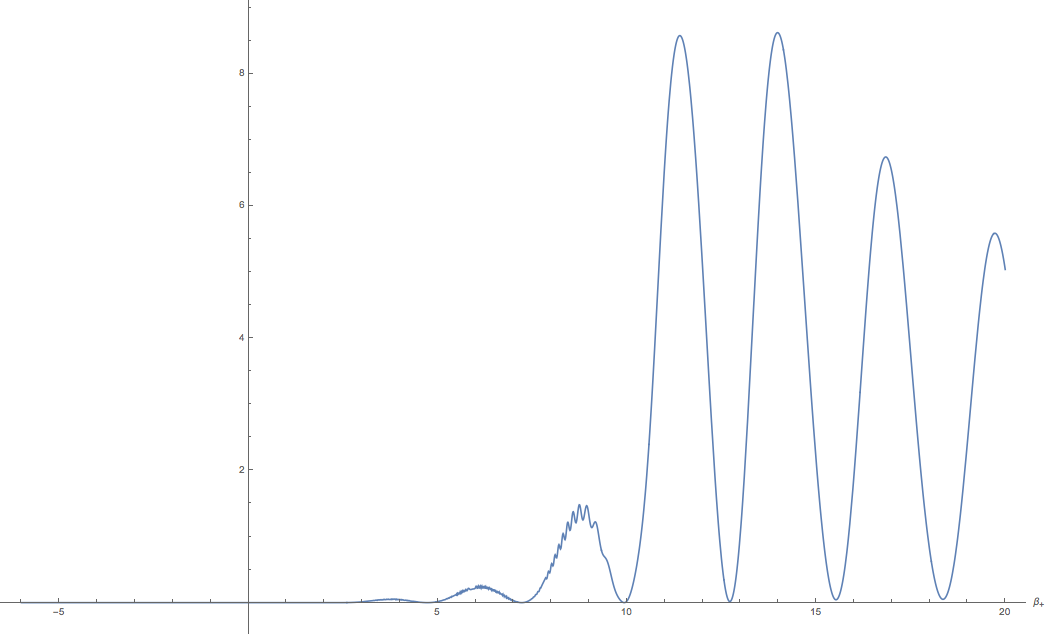}
\caption{$C_{0}$=2 \hspace{1mm} $\alpha$=6}
\end{minipage}%
\end{figure}

\section{\label{sec:level1} Concluding Remarks}

Surprisingly our results for the LRS Bianchi IX and VIII models are qualitatively similar, despite the fact that classically they yield different spatial topologies. A feature of our wave functions is that they possess one large peak centered around some value of the Misner variables with additional smaller peaks around other values. If we assume that larger values of $\abs{\Psi(\alpha,\beta_+)}^{2}$ correspond to greater probabilities that our LRS Bianchi IX or VIII universes will have a geometry dictated by $\left(\alpha,\beta+\right)$; then from our plots above we can say that there is a "preferred" geometry that these universes can possess centered around the tallest peak of our wave functions. In addition the many smaller peaks centered around other values of $\left(\alpha,\beta_+\right)$ are also possible geometries these universes can take on. Furthermore our universes can tunnel between different geometries centered around the many peaks in our wave functions. It should be mentioned though because our Bessel functions are approximately periodic that these "maximal" peaks occur repeatedly as their arguments approach zero. Because our Bessel K function approaches zero for large values of its argument, our solutions  greatly suppress the probability of these universe having geometries where simultaneously $\alpha >> 0$ and $\beta_+ << 0$. Thus as the scale factor related to $\alpha$ increases in the positive direction, our universes will have a tendency to possess a geometry associated with an increasingly positive range of values for $\beta_+$. This can be seen in our 2D plots, as $\alpha$ increases our wave functions travel in the positive $\beta_+$ direction. 

As it can be seen from our 3D plots, the addition of stiff matter causes the peaks around smaller values of $\beta_+$ to be larger than the peaks around larger values of $\beta_+$. For the 2D plots when $\alpha$=-20 it can be seen that stiff matter qualitatively changes how our wave functions look, most notably by shifting the largest peak towards the negative $\beta_+$ direction. Because stiff matter doesn't scale with either $\alpha$ or $\beta_+$ its effects become important when both the Misner variables in the potential of (3) are in a range where the stiff matter term is comparable to rest of the  terms in the potential. 

In conclusion, we have added to the list of closed form solutions to the Wheeler DeWitt equation for an arbitrary Hartle-Hawking parameter. Via superposition we constructed interesting looking wave functions and using an admittingly naive interpretation extrapolated some interesting rough statements on the quantum evolution of LRS Bianchi IX and VIII universes from them. The author looks forward in the future to comparing these results to those obtained by applying the Euclidean-signature semi classical method\cite{marini2019euclidean} \cite{moncrief2014euclidean} to the LRS Bianchi IX and VIII models. 

\section{\label{sec:level1}ACKNOWLEDGMENTS}
 
I am grateful to Professor Vincent Moncrief for valuable discussions at every stage of this work. I would also like to thank George Fleming for facilitating my ongoing research in quantum cosmology. Daniel Berkowitz acknowledges support from the United States Department of Energy through grant number DE-SC0019061. I also must thank my aforementioned parents.

\bibliography{Bianchi}

\end{document}